\documentclass[twocolumn,twoside,slac_two]{revtex4}
\usepackage{graphicx}
\usepackage{fancyhdr}
\usepackage{lscape}
\usepackage{multirow}
\usepackage{longtable}
\begin{document}

\newcommand{\gr}{$\gamma$-ray}
\newcommand{\grs}{$\gamma$-rays}
\newcommand{\vhe}{V\textsc{HE}}
\newcommand{\dg}{\ensuremath{^\circ}}
\newcommand{\nthu}{National Tsing Hua University}
\newcommand{\hess}{H.E.S.S.}

\title{Are GeV and TeV spectra connected?\\
the case of Galactic \gr~sources}

%

\author{P. Tam}
\affiliation{Institute of Astronomy and Department of Physics, National Tsing Hua University, Hsinchu, Taiwan}
\author{S. Wagner}
\affiliation{Landessternwarte, Universit\"at Heidelberg, K\"onigstuhl, Heidelberg, Germany}

\begin{abstract}
To understand Galactic objects that emits GeV-TeV emission, a
spatial correlation study between the Fermi bright source catalog~\cite{bsl_lat} and
TeV source population was carried out in~\cite{Tam10}, finding that a significant number of very high-energy (VHE; E$>$100 GeV) sources are also emitting at GeV energies. We extended our previous study
utilizing the first Fermi catalog (1FGL) sources~\cite{lat_1st_cat}. A cross-correlation
comparison of the 1FGL sources was carried out with the VHE \gr~sources in the
literature as of May 2011. While it is found that a significant number of VHE \gr~sources
were also detected in the GeV band, the GeV-TeV spectra of some of
these spatially coincident sources cannot be described by a single
spectral component. While some of these cases are \gr~pulsars
accompanied by VHE \gr~emitting nebulae, we present cases where
the 100 MeV to multi-TeV spectra of coincident 1FGL/VHE source pairs
do not seem to be well fit by a single spectral component.
\end{abstract}

\maketitle

\thispagestyle{fancy}


\section{Introduction}
During the last decade, many different kinds of astrophysical objects in our Galaxy were discovered at photon energies above 100~MeV: pulsars (PSRs), pulsar wind nebulae (PWNe), supernova remnants (SNRs), high-mass X-ray binaries (HMXBs), and one H~II region.
They were all made by utilizing the high-energy (HE; 30 MeV--100 GeV) and very high-energy (VHE; 100 GeV--100 TeV) \gr~experiments including current generation of imaging atmospheric Cherenkov telescopes (IACTs) H.E.S.S., MAGIC, and VERITAS, and the Large Area Telescope (LAT) aboard the Fermi satellite. 

More than 100 sources are now known at VHE \gr~energies and 1451 sources are listed in the first Fermi LAT catalog, comparing with $\sim$10 VHE \gr~sources and $\sim$300 HE \gr~sources around the turn of the century. Given the large number of sources, we follow previous studies~\cite{Funk08,Tam10} and compare the HE and VHE source positions, as an important step to identify a group of sources emitting both in the HE and VHE bands.


\section{Spatial coincidence study}
We cross-correlated the 1FGL source centroid positions with VHE \gr~source 
centroid positions. Only sources that are not associated with an
extragalactic source were considered. Using the same manner as
described in~\cite{Tam10}, the VHE source extent and 95\% uncertainty in the
1FGL source centroids are taken into account. All first Fermi/LAT catalog 
sources are assumed to be point sources as in~\cite{lat_1st_cat}. Those sources with an ending `c' should be regarded with caution given the imperfect knowledge of the diffuse \gr~background~\cite{lat_1st_cat}. In total we identified
31 1FGL sources that are spatially coincident with one VHE source. In addition, the VHE source in the Westerlund~1 region,
which are $\sim$0.6$\dg$ extended, is found to be spatially coincident
with three 1FGL sources. HESS J1809$-$193 is coincident with two 1FGL sources. The list of these 1FGL-VHE source pairs are
presented in Table~\ref{tab:pos_coincidence1}.

Based on pulsar timing information and dedicated efforts described in the corresponding literature, as well as spatial
coincidences, the 1FGL sources in the list of coincidences include several classes: 2
HMXBs (LS~I~$+$61$\dg$~303 and LS~5039), 8 PSRs, 4 SNRs (IC~443, W28,
W49B, W51C), 2 PSR/PWN (Crab and Vela), 6 SNR/PWN candidates, one
H~II region, and 13 unassiciated sources.

\section{GeV-TeV spectra}

The GeV spectral points are taken from the 1FGL catalog where point source
analysis was used, while the VHE spectra shown are the best-fit
power law taken from the respective literature. 

We identify several
cases of which the 0.1--100~GeV spectra and the VHE spectra cannot be
described by a single spectral components, as shown in
Figs 1--5. The flux in the five energy bands in~\cite{lat_1st_cat} are plotted together
with the best-fit power law in the VHE range. In several other cases, the 
GeV emission come from a \gr~pulsar, i.e., those 1FGL source identified as a pulsar, that shows cutoff at several GeV and VHE emission mostly likely come from
the associated VHE \gr~emitting PWN. We only present cases where the 1FGL source
is not identified as a pulsar.

%


\section{Cases of spectral `mis-match'}
We found five VHE sources that are spatially coincident with a
1FGL source but the GeV--TeV spectra are incompatible with a
single spectral component: HESS~J0852$-$463, HESS~J1614$-$518,
HESS~J1702$-$420, HESS~J1809$-$193, and HESS~J1848$-$018.
The cases presented here might represent a group of GeV/TeV
sources where the spectral mis-matches indicate different
radiations working at different energies or that radiation comes
from different parts of a \gr~source. Further studies of these
spectral mis-match GeV/TeV spatially coincident cases are
encouraged.

\section{Conclusion}
In this study, it is found that a significant number of VHE sources
are spatially coincident with a counterpart in the first Fermi/LAT catalog, establishing a
population of sources that emit both in the HE and VHE energy bands. This confirms our previous assessment using the Fermi bright source list~\cite{Tam10}.

However, the GeV-TeV spectra of some of
these spatially coincident sources cannot be described by a single
spectral component. While some of these cases are \gr~pulsars
accompanied by VHE \gr~emitting nebulae, we highlight five cases where
the 100 MeV to multi-TeV spectra of coincident 1FGL/VHE source pairs
do not seem to be well fit by a single spectral component.

\emph{Notes added in proof}: The second Fermi catalog has been released after the conference. We note that one of the coincidence pairs, 1FGL~J1702.4$-$4147c, does not have a 2FGL counterpart.

%


\begin{table*}
\begin{minipage}[t]{180mm}
{\scriptsize
\caption{1FGL sources with spatially coincident VHE counterpart as of May 2011. The class denoted `SNR/PWN' means SNR/PWN candidates, according to~\cite{lat_1st_cat}.}
\label{tab:pos_coincidence1}
\centering
\renewcommand{\footnoterule}{}  
\begin{tabular}{lccrr|c@{}crrcc}
\hline\hline
     1FGL source       & association & class & $l$ & $b$& VHE $\gamma$-ray source & association & $l$ & $b$ & extension & ref \\
               &  & & (\dg) & (\dg) & & & (\dg) & (\dg) & (\dg) & \\
    \hline
    J0240.5$+$6113 &LS~I$+$61~303 &HMXB &135.66 &1.08 &VER~J0240$+$612 &LS~I~$+$61~303 &135.70 &1.08 &pt src  &\cite{veritas_LSI61} \\
    J0534.5$+$2200 &Crab    &PSR/PWN &184.56 &$-$5.76 &HESS~J0534$+$220 &Crab nebula &184.56 &$-$5.78 &pt src  & \cite{hess_crab}\\
    J0617.2$+$2233 &IC~443 &SNR &189.08 &3.07  & VER~J0616.9$+$2230 & IC~443 &189.08 &2.92 & 0.16  & \cite{lat_ic443,veritas_ic443}\\
    J0835.3$-$4510 &Vela    &PSR/PWN &263.56 & $-$2.77 &HESS~J0835$-$455 & Vela~X & 263.86 & $-$3.09 & 0.43  &\cite{hess_velaX} \\
    J0854.0$-$4632 &    &SNR/PWN &266.64 & $-$1.09 &HESS~J0852$-$463 &RX~J0852.0$-$4622 &266.28 &$-$1.24 & 1.0  &\cite{hess_snr0852} \\
    J1023.0$-$5746 &PSR~J1023$-$5746 &PSR &284.17 &$-$0.41 & HESS~J1023$-$575 &PSR~J1023$-$5746/Wd~2 &284.22 &$-$0.40 &0.18  &\cite{hess_westerlund2_1023_paper2} \\
    J1418.7$-$6057 &PSR~J1418$-$6058 &PSR &313.34 &0.11 &HESS~J1418$-$609 &G313.3$+$0.1 (Rabbit) &313.25 &0.15 & 0.06  &\cite{hess_kookaburra} \\
    J1420.1$-$6048 &PSR~J1420$-$6048 &PSR &313.50 &0.20 &HESS~J1420$-$607 &PSR~J1420$-$6048 &313.56 &0.27 &0.07  &\cite{hess_kookaburra} \\
    J1501.6$-$4204	&        &SNR/PWN &327.30 &14.54  &HESS~J1502$-$421 &SN~1006~SW &327.35 &14.48 &0.13  & \cite{hess_sn1006} \\
    J1503.4$-$5805c &        &Unid    &319.67 &0.42   &HESS~J1503$-$582 &FVW~319.8$+$0.3? &319.62 &0.29 & 0.26  &\cite{hess_1503_fvw} \\
    J1614.7$-$5138c &        &Unid &331.69 &$-$0.49 &HESS~J1614$-$518 & &331.52 &$-$0.58 &0.2  &\cite{hess_survey} \\
    J1640.8$-$4634c &       &SNR/PWN &338.29 &$-$0.06 &HESS~J1640$-$465 &G338.3$-$0.0 &338.32 &$-$0.02 &0.05  &\cite{hess_survey} \\
    J1648.4$-$4609c &PSR~J1648-4611 &PSR &339.47 &$-$0.79 &Westerlund 1 region & &339.55 &$-$0.40 &$\sim$0.9  &\cite{hess_westerlund1_1648} \\
    J1649.3$-$4501c &        &Unid &340.44 &$-$0.18 & same as above & & & & & \\
    J1651.5$-$4602c &        &Unid &339.91 &$-$1.12 & same as above & & & & & \\
    J1702.4$-$4147c &        &Unid &	344.45  &0.00 &HESS~J1702$-$420 &PSR~J1702$-$4128 &344.26 &$-$0.22 &0.3  &\cite{hess_dark} \\
    J1707.9$-$4110c &        &Unid &	345.56  &$-$0.44 &HESS~J1708$-$410 & & 345.67 &$-$0.44 &0.08  &\cite{hess_dark} \\
    J1709.7$-$4429 &PSR~B1706$-$44 &PSR &343.10  &$-$2.69  &HESS~J1708$-$443 &PSR~B1706$-$44/G343.1$-$2.3 &343.06 &$-$2.38 &0.29  &\cite{hess_psr1706}\\
    J1711.7$-$3944c &        &SNR/PWN &	347.15  &$-$0.19 &HESS~J1713$-$397 &RX~J1713.7$-$3946 &347.28 & $-$0.38 &0.25  &\cite{hess_snr1713} \\
    J1718.2$-$3825 &PSR~J1718$-$3825 &PSR &349.00 &$-$0.40 &HESS~J1718$-$385 &PSR~J1718$-$3825? &348.83 &$-$0.49 & 0.015  &\cite{hess_psr_1718_1809} \\
    J1745.6$-$2900c & &SNR/PWN &359.94 &$-$0.05 &HESS~J1745$-$290 &Sgr~A*/G359.95$-$0.04 &359.94 &$-$0.04 &pt src &\cite{hess_GC_pos} \\
    J1800.5$-$2359c &W28$-$A2 & H~II region&5.95 &$-$0.37 &HESS~J1800$-$240B &W28$-$A2 &5.90 &$-$0.37 &0.15  &\cite{lat_w28,hess_1801-233}\\
    J1801.3$-$2322c &W28 &SNR &6.57 &$-$0.22 &HESS~J1801$-$233 &W28 &6.66 &$-$0.27 &0.17  &\cite{lat_w28,hess_1801-233}\\
    J1805.2$-$2137c &        &SNR/PWN &8.55 &$-$0.14 &HESS~J1804$-$216 &W30/PSR~J1803$-$2137? &8.40 &$-$0.03 &0.20  &\cite{hess_survey} \\
    J1808.5$-$1954c &      &Unid  &10.43 &0.03 &HESS~J1809$-$193 &PSR~J1809$-$1917? &10.92 &0.08 &0.53  &\cite{hess_psr_1718_1809} \\
    J1810.9$-$1905c &      &Unid  &11.42 &$-$0.08 & same as above & & & & & \\
    J1826.2$-$1450	&LS~5039  &HMXB  &16.88 &$-$1.29  &HESS~J1826$-$148 &LS~5039 &16.90 &$-$1.28 &pt src  &\cite{hess_LS5039} \\
    J1837.5$-$0659c & &Unid &25.13 &$-$0.12 &HESS~J1837$-$069 & & 25.18 & $-$0.12 & 7.2'$\times$3'  &\cite{hess_survey} \\
    J1844.3$-$0309c &       &Unid  &29.32 &0.13 &HESS~J1843$-$033 &&$\sim$29.08 & $\sim$0.16 &$\sim$0.2  &\cite{hess_survey_2007} \\
    J1848.1$-$0145c &       &Unid  &30.99 &$-$0.08 &HESS~J1848$-$018 &&30.98 &$-$0.16 &0.32  &\cite{hess_1848_HDGS}\\
    J1907.9$+$0602 &PSR~J1907$+$0602&PSR &40.18&$-$0.89&HESS~J1908$+$063&MGRO~J1908$+$06 &40.39 &$-$0.79 &0.34 &\cite{hess_1908+063} \\
    J1910.9$+$0906c &W~49B &SNR &43.25 &$-$0.16 &W~49B~region &W49B & 43.26 &$-$0.19 &pt src  &\cite{lat_w49b,hess_w49} \\
    J1913.7$+$1007c &       &Unid  &44.48 &$-$0.28 &HESS~J1912$+$101 & PSR~J1913$+$101 &44.36 &$-$0.08 &0.27  &\cite{hess_psr1913}\\
    J1922.9$+$1411	&W~51C &SNR &49.12 &$-$0.38 &HESS~J1923$+$141 &W51 & 49.14 &$-$0.6 &$\sim$0.15  &\cite{lat_W51C,hess_w51}\\
    J2020.0$+$4049	&        &Unid  &78.37 &2.53 &VER~J2019$+$407 & $\gamma$ Cygni SNR? & 78.33 & 2.54 & 0.16$\times$0.11  &\cite{VERITAS_cygnus_survey09} \\
    J2032.2$+$4127 & PSR~J2032.2$+$4127 &PSR &80.22 &1.03 &TeV~J2032$+$4130 &&80.23 &1.10 &0.10  &\cite{hegra_TeV2032}\\
    \hline
    \end{tabular}
}
\end{minipage}
\end{table*}

\begin{figure*}
\includegraphics[width=85mm]{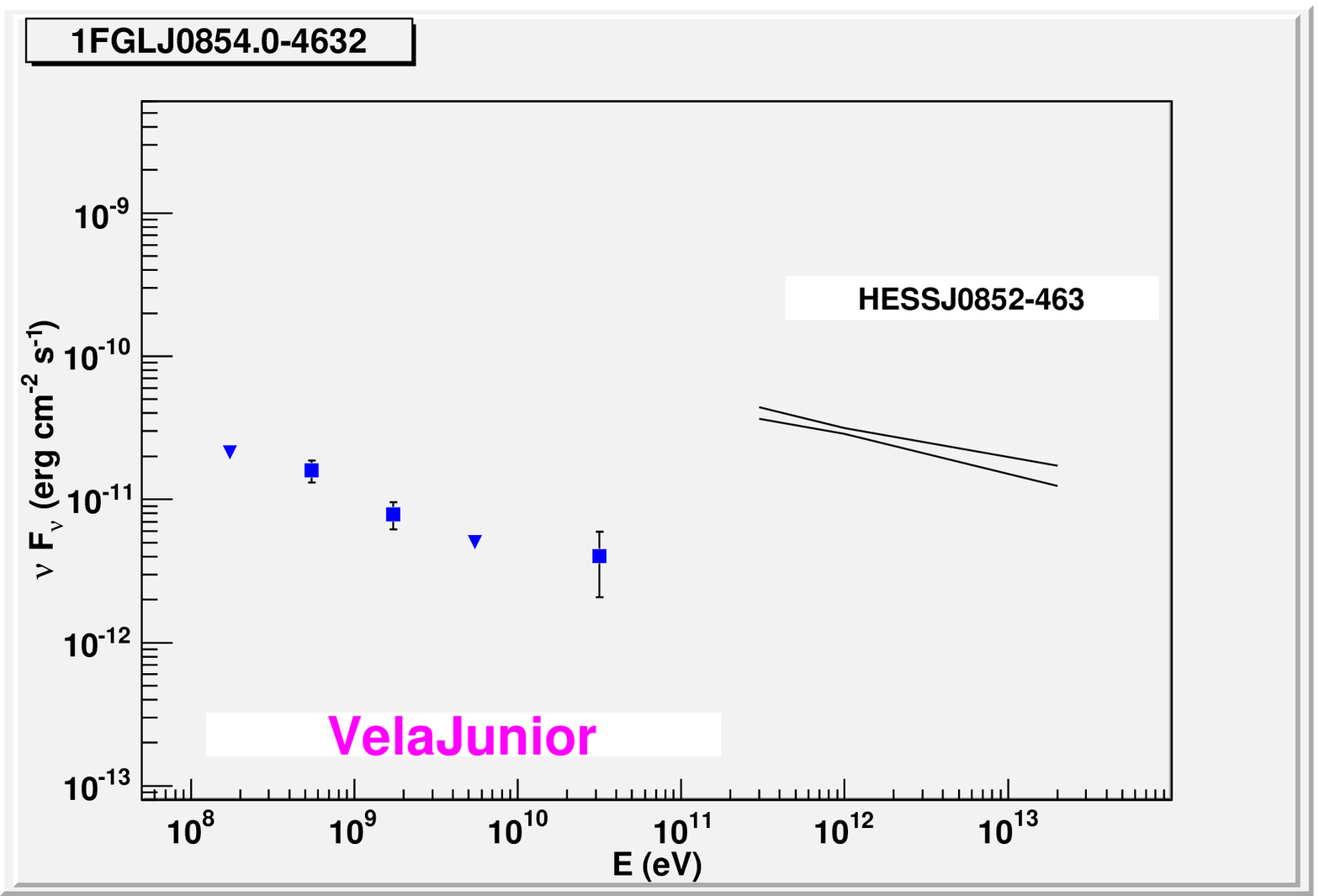}%
\includegraphics[width=85mm]{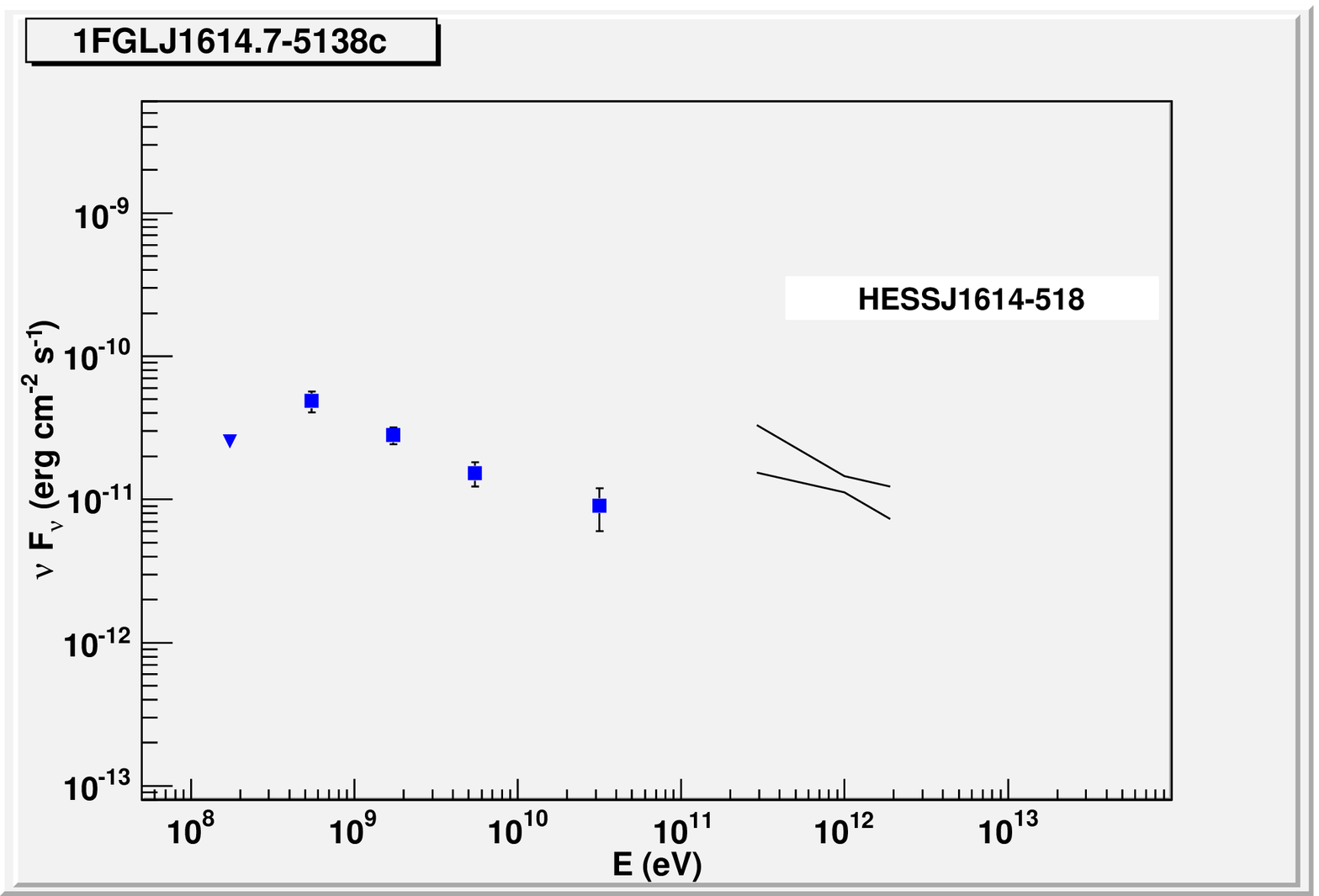}\\%
\includegraphics[width=85mm]{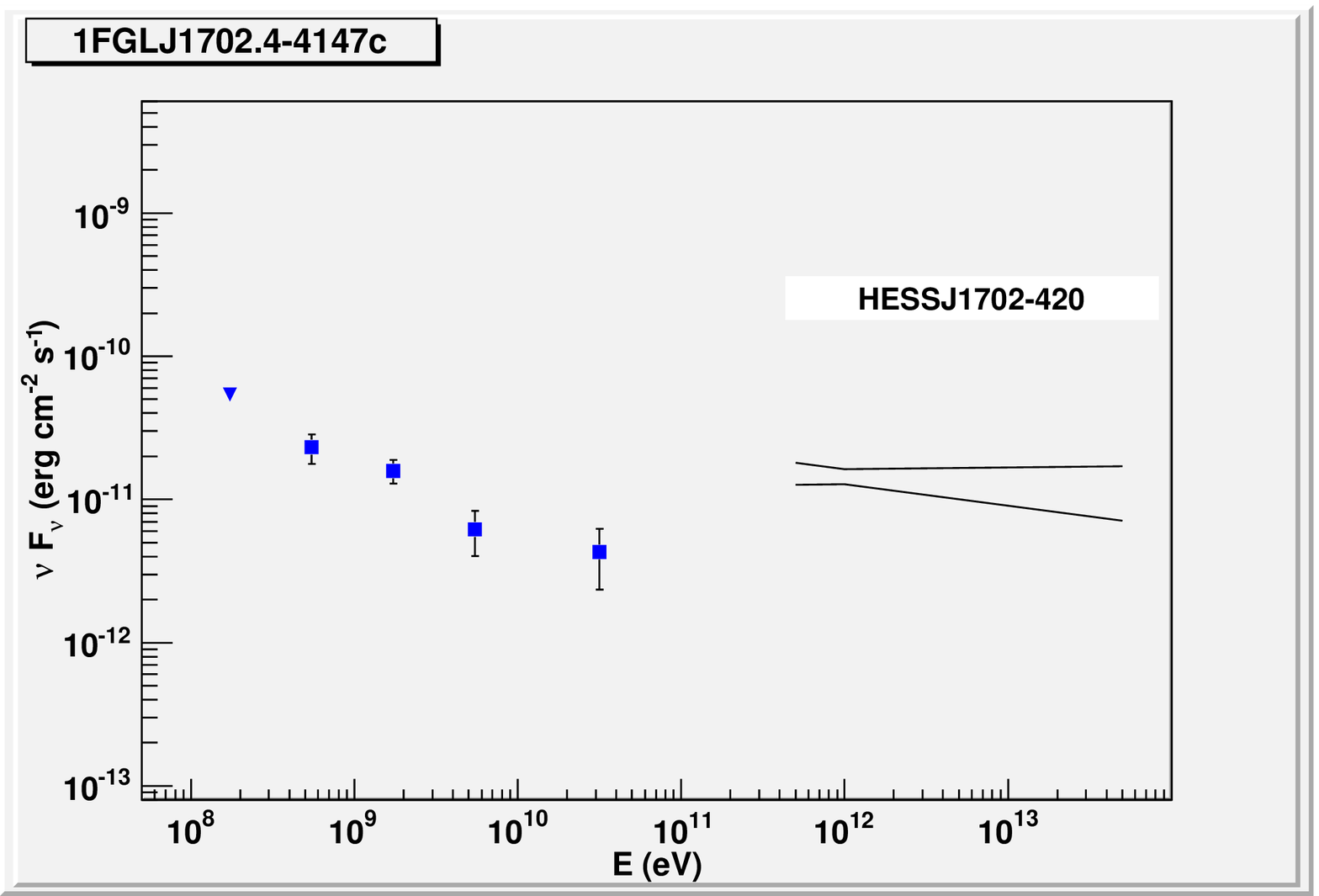}%
\includegraphics[width=85mm]{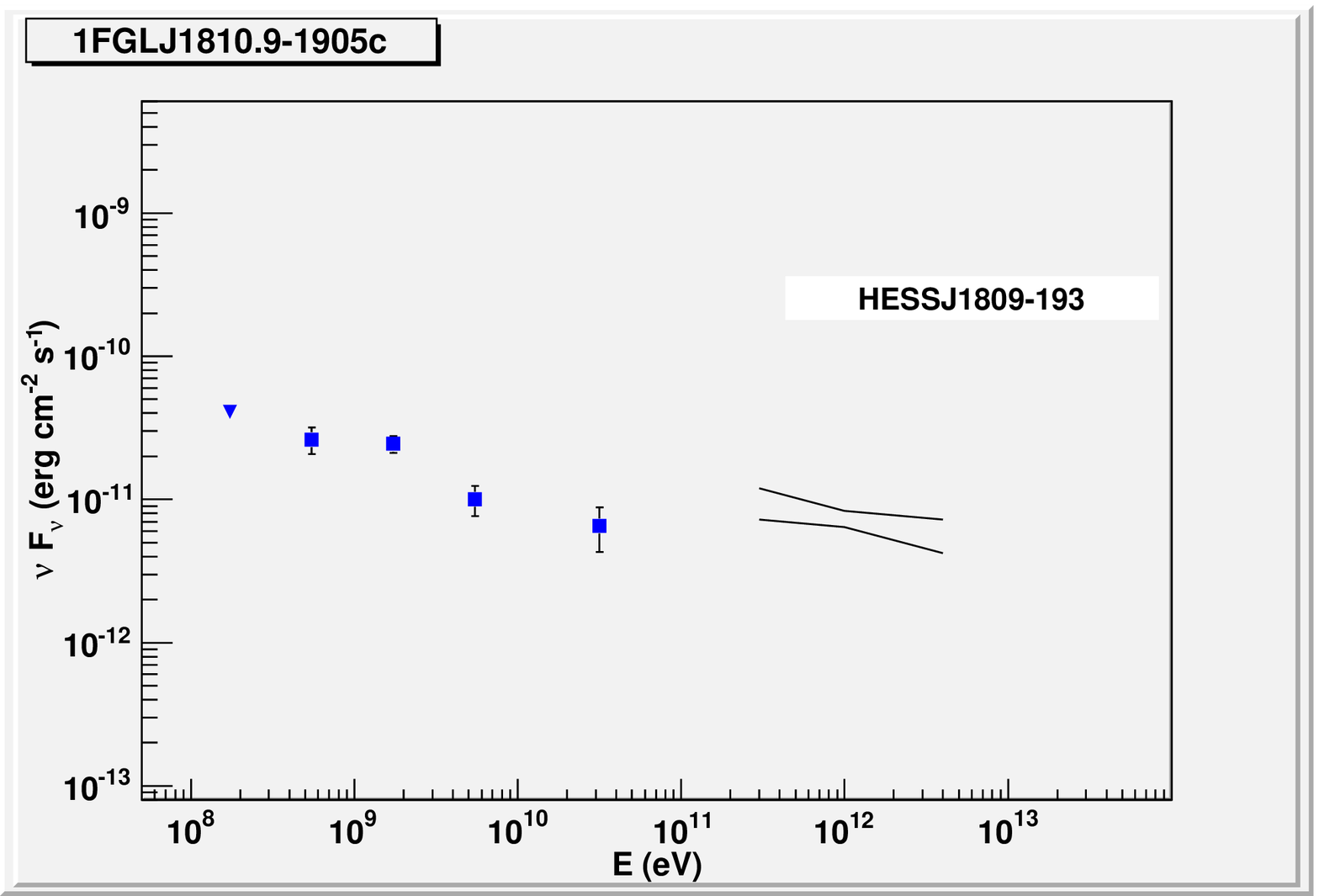}%
\caption{The 100~MeV to several tens TeV spectra of four spatially coincident but spectrally `mis-match' 1FGL/VHE source pairs}
\end{figure*}

\begin{figure*}
\includegraphics[width=85mm]{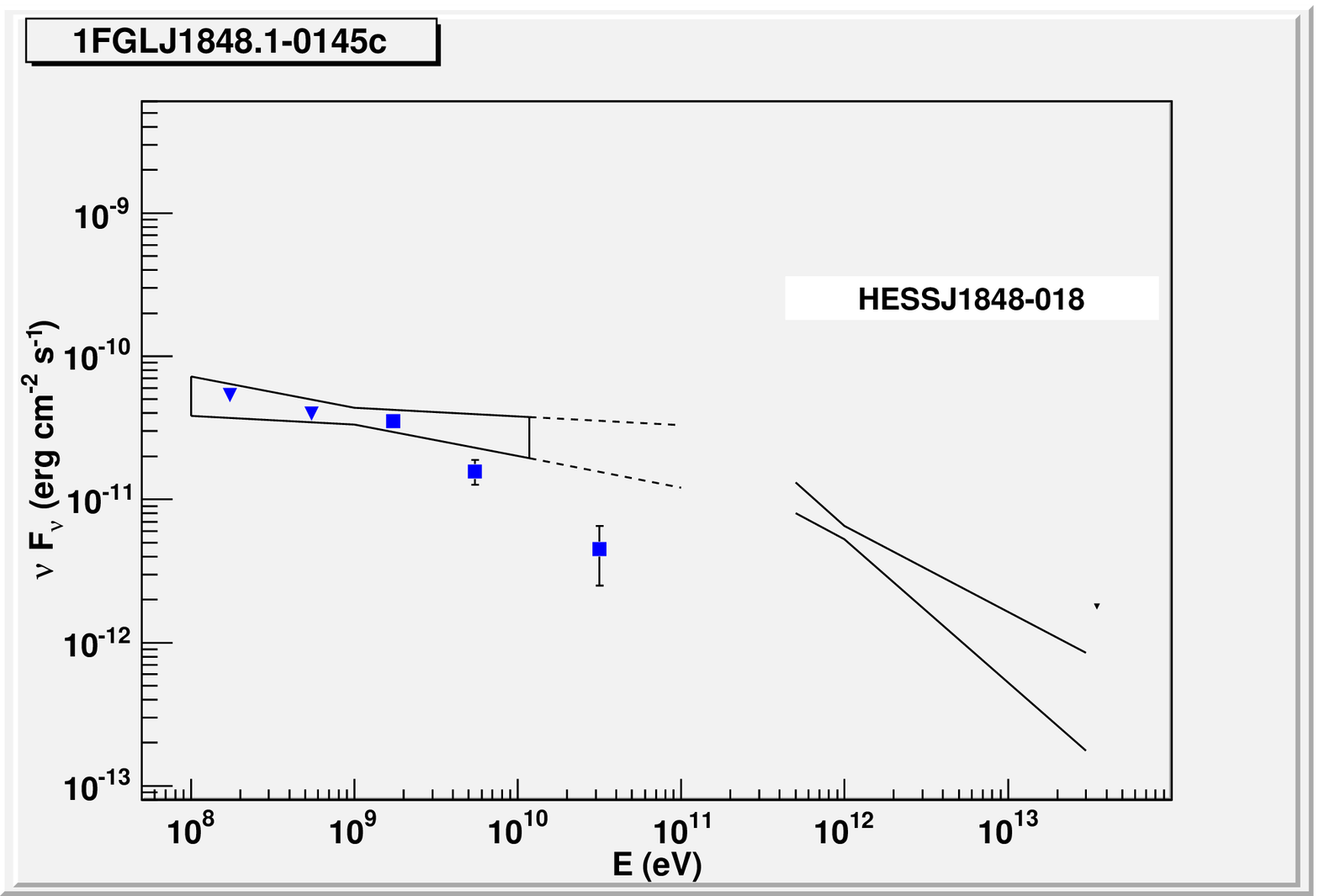}
\caption{The 100~MeV to several tens TeV spectra of 1FGL~J1848.1-0145c and HESS J1848-018.
1FGL~J1848.1-0145c has a 0FGL counterpart whose best-fit power law is also shown.}
\end{figure*}




\bigskip 
\begin{acknowledgments}
P. Tam acknowledges the support of the Formosa Program of Taiwan, NSC100-2923-M-007-001-MY3, and
the NSC grant, NSC100-2628-M-007-002-MY3.

\end{acknowledgments}

\bigskip 

\end{document}